Paper ID 576

# Cross-Platform Simulation Architecture with application to truck platooning impact assessment

**Andres Ladino[1*], Lin Xiao[2], Kingsley Adjenugwhure[2], Nicolás Deschle[2], Gerdien Klunder[2]**

1. andres.ladino@univ-eiffel.fr , LICIT, Université Gustave Eiffel, France
2. {lin.xiao, kingsley.adjenughwure, nico.deschle, gerdien.klunder}@tno.nl, TNO, Netherlands

**Abstract**

Simulation-based traffic impact assessment studies of advanced technologies such as truck platooning need to be carried out to ascertain their benefits for traffic efficiency, safety and environment. To reduce uncertainty in the results of such simulation-based studies, the same simulation studies can be performed in different simulation software. Many traffic simulation software packages (Aimsun, SymuVia, Vissim, SUMO) are currently available for traffic impact assessment of new technologies such as truck-platooning. However, to fully model and simulate the functionalities of such advanced technologies in different simulation environments, several extensions need to be made to the simulation platforms. In most cases, these extensions have to be programmed in different programming languages (C++, Python) and each simulator has its own simulator specific API. This makes it difficult to reuse software written for a specific functionality in one simulation platform in a different simulation platform. To overcome this issue, this paper presents a novel architecture for cross-platform simulation. The architecture is designed such that a specific functionality such as truck-platooning or any other functionality is made platform independent. We designed a cross-platform architecture for simulating a truck-platooning functionality using Vissim and SymuVia simulation software to determine the traffic flow effects of multi-brand truck platooning in the context of the EU project ENSEMBLE. In this draft paper, we present the structure of the framework as well as some preliminary results from a simple simulation performed with the cross-platform simulator.

**Introduction**

Recent advances on truck automation have shown to be the promising future towards full automated driving. The ENSEMBLE project paves the way into the development of these new technologies by bringing truck platooning to a multiple brand framework feasible and deployable. Inherently, platoon technologies have a number of consequences in traffic and the environment and its impact needs to be evaluated. Among the different approaches, traffic simulations by means of microsimulation software is a frequent choice due to its versatility, relative cost and scalability compared to experiments.

Simulation Framework for Multi-brand Truck Platooning in Traffic Networks

To assess the performance of the technology, more than one simulation platform serves to reduce impact of specific models implemented independently. Also, different simulators can be chosen based on different strengths and convenience to simulate different kind of networks and properties. For such multi-platform simulation studies, a cross-platform design of the functionality is needed for two reasons: firstly, to provide an open platform for model development and secondly, to ensure that differences in results due to simulation platform specific implementations of the technology are greatly minimized.

In this paper, the implementation of a cross-platform architecture tailored to classic traffic micro simulators is presented. The cross-platform architecture is designed to offer flexibility in terms of collecting and sending data from and to the different simulators together with a platform independent implementation of the functionality (in this case truck platooning). In particular, the full architecture connects two traffic simulators, Vissim[1] and SymuVia[2], to a common application interface in order to simulate interactions between platoons and regular vehicles. The logic of the system is divided into multiple logical layers facilitating the decision process when platoons interact with regular traffic while respecting constraints of the cyber physical system.

**Truck Platooning**

Platooning technology has made significant advances in the last decade, but to achieve the next step towards deployment of truck platooning, an integral multi-brand approach is required. Aiming for Europe-wide deployment of platooning, 'multi-brand' solutions are paramount. It is the ambition of ENSEMBLE to elaborate pre-standards for interoperability between trucks, platoons and logistics solution providers, to speed up actual market pick-up of (sub)system development and implementation and to enable harmonization of legal frameworks in the member states. The main goal of the ENSEMBLE project is to pave the way for the adoption of multi-brand truck platooning in Europe to improve fuel economy, traffic safety and throughput.

Improvements and impacts introduced by truck platooning are assessed as part of the project by considering several aspects such as road infrastructure, economic and environmental benefits, behavior of truck drivers and other road users and finally traffic flow. Special focus is assigned to the yet-unexplored Multi-brand case where the impact assessment particularly challenging due to the heterogeneity in truck characteristics i.e., different load, different braking capabilities and the use of unknown ('black box') platooning algorithms (these are usually kept confidential by each brand), which makes more difficult to ensure platoon stability. For the road infrastructure case, variability on load and dimensions of the platoons are examined in order to determine potential consequences on the infrastructure. In the case of economic and environmental benefits, the main idea is to understand and design economical models to determine and generate added value for platooning strategies. These models may imply effects of formation of platoons on the fly or hub-based formations. At the same time the objective is understanding the impact of assisted driving and cooperative technologies on drivers and finally to determine the achievable effects of truck platooning under real world conditions on roads in the presence of specific traffic conditions, such as overtaking traffic, congestion, and merging vehicles from on- and off ramps.

---

[1] Vissim is a commercial platform simulator developed by PVT Group and used by TNO. See more information at https://www.ptvgroup.com/en/solutions/products/ptv-Vissim/

[2] SymuVia is an open-source microscopic dynamic traffic simulator developed by UGE. See more information at https://github.com/Ifsttar/Open-SymuVia

Some former works have presented traffic impact of platoons in highway traffic operations and traffic flow. In (Kunze et al. 2009), initial definitions have been provided in order to deploy platooning operations in regular highways. Effects of such implementations were studied in (Calvert, Schakel, and van Lint 2017), where it was found that potential effects of automated technologies may be observed only for high penetration rates. (Cicic and Johansson 2019b; 2019a) have detailed and exploited macroscopic models to explain the moving bottleneck effects that platoons may induce in traffic flow. Although these models are of high computing efficiency, specific effects regarding vehicle interactions may produce instabilities that are not defined for such models. (Calvert, Schakel, and van Arem 2019) performed detailed impact traffic assessment by developing quantitative proof of the potential effects of truck platooning on traffic flow performance. The results showed that truck platooning may have a small negative effect on the total non-saturated traffic flow, however with a much larger negative effect on saturated traffic flow. More recently (Jin et al. 2020) proposed tandem-link fluid models that consider randomly arriving platoons sharing highway with regular vehicles by defining control architectures and their impacts in such situations. These works have settled an initial base that considers platoon specifications complementary to the ENSEMBLE project. In our case, the multi-brand heterogeneous factor and a detailed hierarchical logical layer have been designed to model accurate interactions between platoons and regular traffic.

The special focus on impact on traffic flow requires to examine and measure behaviour of traffic conditions surrounding a truck platoon, such as the individual variability of headway space, speed and acceleration of a formation of trucks in a heterogeneous platoon and the total time taken to perform specific manoeuvres in traffic. These performance indicators allow for better understanding and comprehension of the platoon operation under regular traffic conditions.

**Framework of two connected simulators**

The proposed application is designed as a command line interface where commands are interpreted and then transmitted towards traffic simulators via a standard socket communication scheme. A scheme of the full framework can be observed in Figure 5. In order to connect both traffic simulators we consider SymuVia and Vissim via a Shared Library[3] in the former case and a port communication in the later

---

[3] Shared Libraries are compiled binary files. Dynamic Link Library (DLL) corresponds to a specific implementation of such binary files for Microsoft ® platforms.

Simulation Framework for Multi-brand Truck Platooning in Traffic Networks

one. The main objective of these sockets is to transmit information and commands towards the corresponding platform such that the dynamic of the full traffic is evolved according to the platoon definitions.

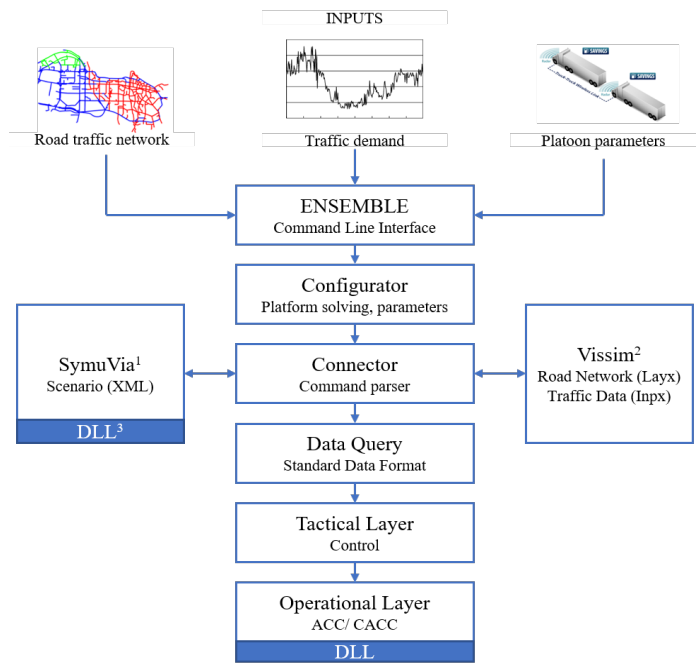

**Figure 1. Proposed cross-platform framework to emulate truck platoons in connection with traffic simulators**

Figure 1 illustrates the general workflow of information starting from the input to the system until it reaches the specific automated vehicles. The application receives three main data inputs namely, a representation of the traffic network (XML for SymuVia, LAYX for Vissim); information regarding the traffic behaviour, in particular, traffic demand for the incoming nodes; and finally, a parametrization of the platoon in terms of vehicle generation and origin destination.

The application processes the data via a configurator, which is in charge of verifying the integrity of the input data and afterwards it starts the evolution of the full traffic simulator by enabling the corresponding platforms and connectors. A standard data format is used to describe vehicle information and software patterns are used for the implementation of such blocks to guarantee efficiency and data integrity protocols in the general framework. Traffic state is pulled from the simulator via a Data Query function which afterwards is sent to the tactical layer. This block considers specific vehicle types to define platoons in the simulation and assign the corresponding behaviour at macro level. Tactical decisions involve general manoeuvres like platoon splits, reaction to cut-ins, or beginning of a formation.

*Tactical layer*

The ENSEMBLE project has considered the platooning function as a support functionality for drivers at a longitudinal level. General manoeuvres are considered in use cases handled by a state machine in charge of determining low level information that should be transmitted to each of the trucks. Former works (Duret, Wang, and Ladino 2018) have used this type of approaches to conduct optimal merging. In this case the objective of the tactical layer is to provide an interface to build up the communication layer that will send information through the whole platoon. Two main streams are considered; first, the rear gap coordinator collects platoon information of the ego truck and information upstream of this vehicle such as the maximum length of the platoon, and the position in the platoon. The second stream receives data downstream of the ego truck and collects information related to platoon vehicles downstream of the ego truck, in particular information about the immediate leader. These two blocks determine the ego truck state which can be classified as a discrete set (StandAlone, Joining, Platooning, FrontSplit, BackSplit, Cut-in) defining steady state or transition actions the operational control should perform for the current ego truck (see Figure 2). Once the state is determined, specific transitions are defined among those states via a logic state machine. This design aims to ensure safety in the manoeuvring as well as smooth dynamic transitions operated by the operational layer.

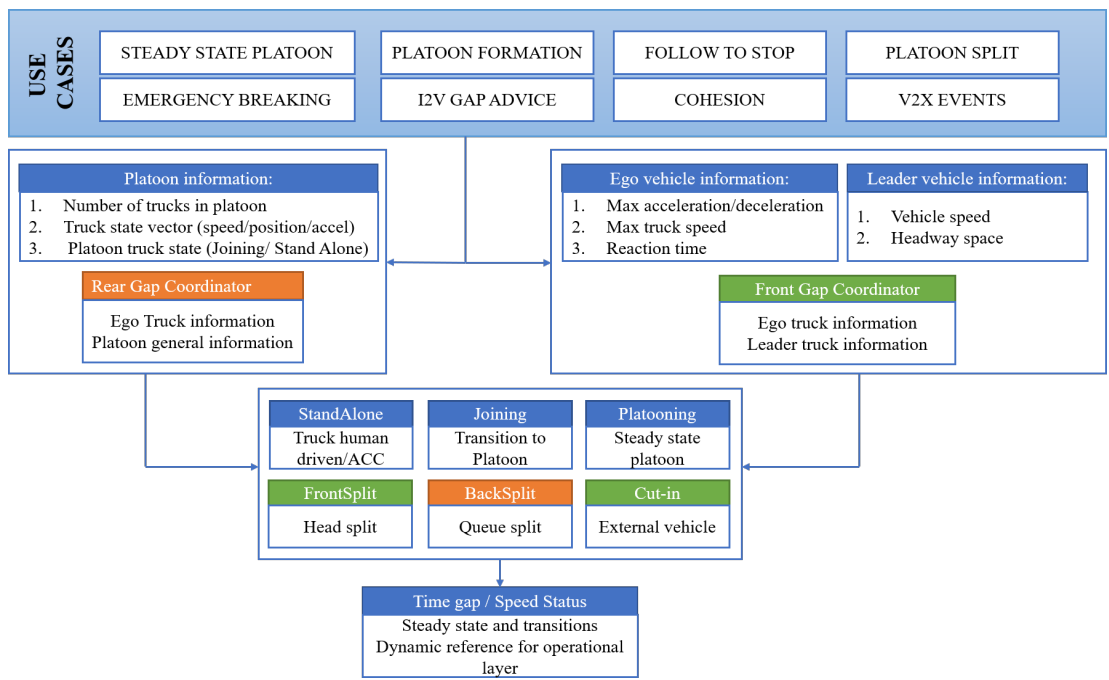

Figure 2. Tactical layer framework

*Operational layer*

Decisions taken by the state machine defined in the tactical layer are transformed into time gap and speed values that are recovered as an entry point for the operational layer, which interconnects high-level decisions from the general platoon with low level control decisions operated by the Adaptive Cruise Control (ACC) and its cooperative counterpart (CACC) when platoons are created. Several works as (Wang et al. 2018) have defined stability conditions for the formation in order to be string stable. In our case the heterogeneity is considered by modelling specific aspects of the truck dynamics such as acceleration bounds or vehicle speed distribution. These quantities depend on both truck static

Simulation Framework for Multi-brand Truck Platooning in Traffic Networks

parameters such as weight, load, engine power and deceleration capabilities and truck state variables such as previous speed and gear.

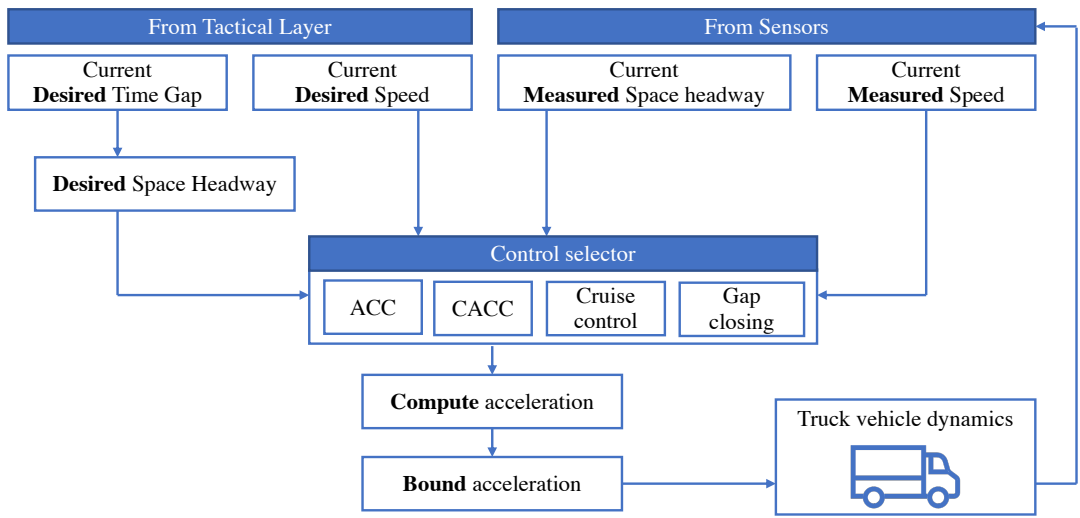

**Figure 3. Operational layer framework**

Figure 3 illustrates the specific workflow implemented in this case. Reference time gaps coming from the tactical layer (setting of the ACC, typically 1.6s) are transformed into desired space headways, the desired speed is also transmitted to the local controller. Virtual sensors enabled in each truck measure speed and estimate the current headway space. By combining the desired values with the current ones, the local ACC computes the control acceleration that should be applied to the truck. A subspace of accelerations is considered by fixing boundaries depending on the vehicle brand or specific vehicle parameters. The loop is then closed by applying the acceleration and evolving the truck dynamics. This stage as explained later is pushed into the traffic simulator, so the evolution of non-platoon vehicles is also computed.

*Authority transition*

An authority transition block is designed to determine whether the vehicle will be controlled by automated systems or a human driver in the next time step, by evaluating the driving environment and vehicle/system performance. It allows a switch between automated driving and manual driving in a simulated automated vehicle, mimicking the real operation of automated vehicles in a mixed traffic environment especially at the low levels of driving automation. Based on the complex decision-making process, a multi-layer architecture is proposed (illustrated in Figure 4), integrating the various authority transition classifications, driving scenarios, driver decision/willingness and driver behavioural assumptions.

Simulation Framework for Multi-brand Truck Platooning in Traffic Networks

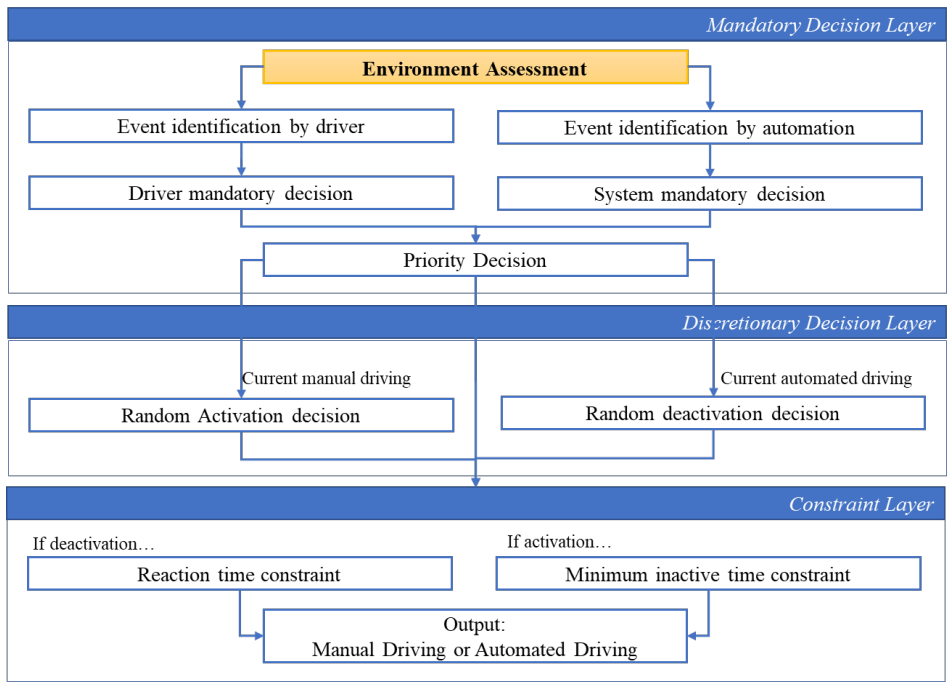

**Figure 4. Authority transition framework**

The Authority transition model is designed as a decision model, generating the decision of "manual control" or "automated control" at each time step. Mandatory decisions and discretionary (optional) decisions are structured as two different layers of the decision-making and a transition constraint layer is designed to model the delay between generating a decision and the realization.

- The authority transition model starts with the mandatory decision layer, where a decision has to be made as a response to certain events/driving situations. Two parallel paths are designed, being driver-initiated decision path and system-initiated decision path. Each path goes through an event identification block and a decision block and the decisions from two paths are prioritized in a final priority decision block.
- The second layer is a discretionary decision layer, where decisions are made by the driver's will instead of by the driving environment. In any moment, a driver is able to activate or deactivate the automated driving system, if no mandatory decision has to be made. The discretionary decision is modelled as a random event.
- A layer of transition constraints is designed as a bottom layer in the architecture, modelling the time delay from decision to realization. A reaction time is assumed when the vehicle control is passed from automated driving systems to human driving and a minimum inactive time is assumed after the driver takes over.

*Integration*

Finally, the integration layer is presented in Figure 5, where the full blocks are presented in an interconnected way. First, the interface performs a request of data to the simulator by asking the current traffic state at time *t*. This step will generate vehicles, vehicle routing and traffic assignment according

Simulation Framework for Multi-brand Truck Platooning in Traffic Networks

to the demand profile, providing the current vehicle state for all vehicles in the simulation. Once information is provided, an assessment environment determines the activation of platoons for ego trucks suitable and desiring to form a platoon. This aspect regulates the transition between the regular human driven model, or the platoon closed loop shown in Figure 3 and Figure 4. Information from the environment is transmitted to the tactical layer via the Front and Rear Gap coordinators as well as external perturbations such as cut-in vehicles. This information creates a use case that determines the decision process of the tactical layer. Low level control is performed updating states of platoon-controlled vehicles. This information is pushed back to the simulator so the traffic platform may compute the dynamic evolution of other vehicles currently present in the network. Once this process is completed, the full cycle starts again in a dynamic evolution evolving a single time step. The framework is able to handle specific events that can be sent to the vehicles such as an emergency stop or a manual condition for drivers to leave the platoon. In this case such events interrupt the platooning process by imposing the human driven behaviour.

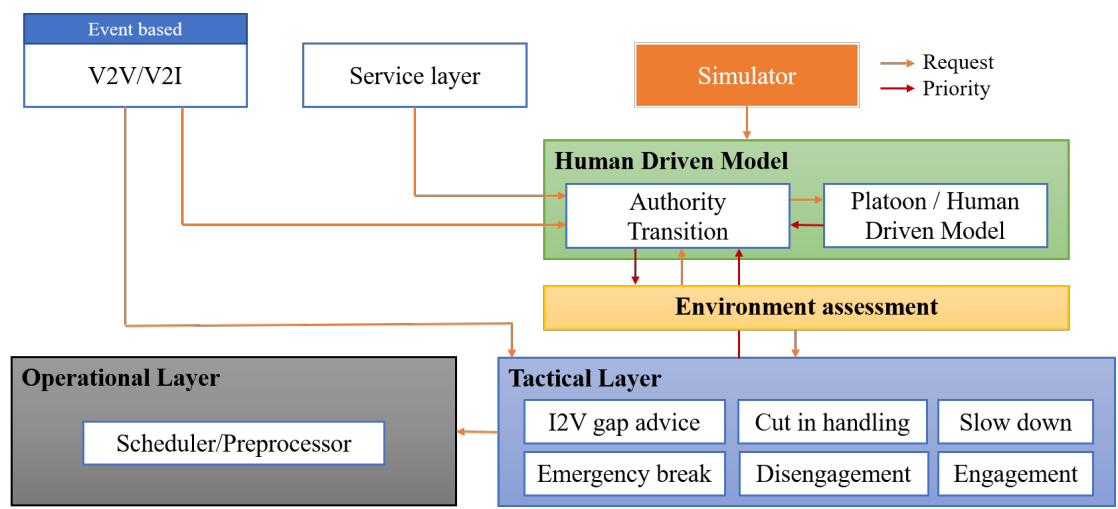

**Figure 5. Integration of the full framework**

**Scenarios and results**

As explained before, this simulation framework is used for the purpose of a simulation study for the ENSEMBLE project to determine the traffic flow effects of multi-brand truck platooning. A number of scenarios has been proposed to evaluate these effects under different circumstances. Some specific setups and situations are shown in Figures 6 and 7 to illustrate the operational behavior, including the cases of stop and go, and cut-in.

The stop-and-go represents the case in which the leader of a platoon composed by three vehicles moving at cruise speed starts decelerating with constant deceleration until stopping. After a period of being standing still it accelerates with a constant acceleration of the same magnitude used to decelerate until reaching cruise speed again. A particular example for cruise speed $v = 32\ m/s$ and acceleration of $a = g/80$ is shown in Figure 6 (a, b, c) where the position, velocity and acceleration for three vehicles in platoon are shown. Leader, second and third followers are denoted on red, blue and green, respectively.

Simulation Framework for Multi-brand Truck Platooning in Traffic Networks

The followers adequately react to the leader without crashing nor creating large gaps. It can be seen in Figure 6 (d), the headway becomes very small when the leader stops but recovers quickly after.

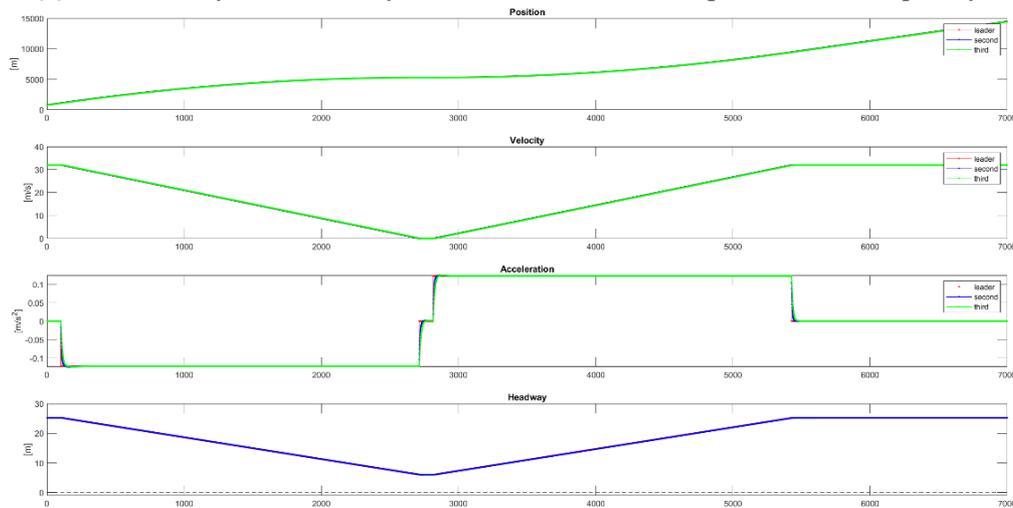

Figure 6. Simulation of the operational layer on the case of a stop-and-go. In (a), (b), (c) the position, velocity and acceleration is shown. Red, blue and green indicate the first three vehicles in the platoon. (d) Shows the headway between the second vehicle in the platoon to its leader, in this case, the leader of the platoon.

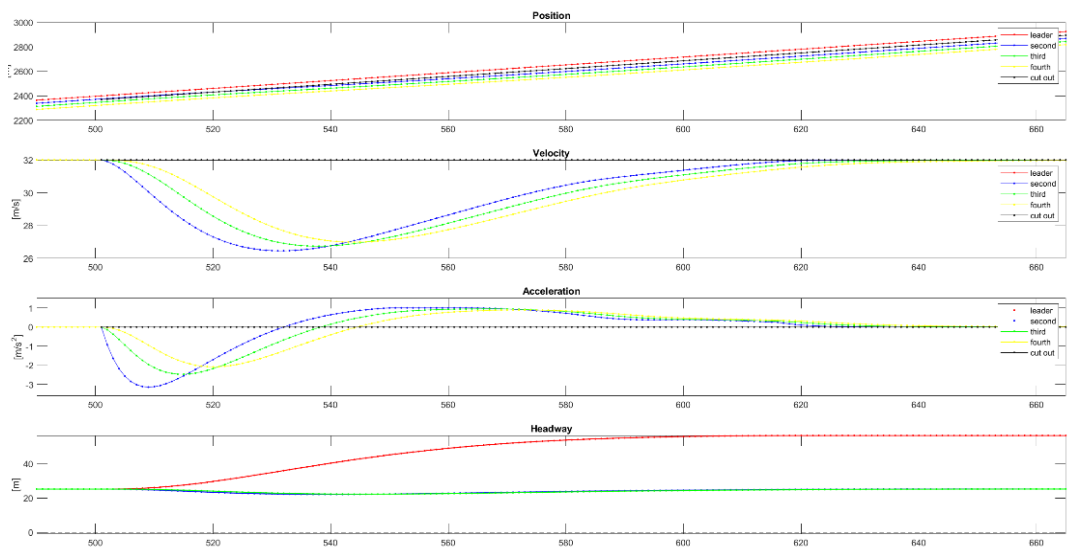

Figure 7. Detail of the simulation of the operational layer on a cut-in. In (a), (b), (c) the position, velocity and acceleration is shown. Red, blue, green and yellow indicate the first four vehicles in the platoon, black denotes the cut-in vehicle.

An important case in a platoon simulation is the one in which a new vehicle occupies the space between two vehicles in a platoon. Figure 7 shows the detail of the behavior of a platoon over the $20\ s$ in which a new vehicle (depicted in black) takes the second position in the platoon. In red, green, blue and yellow, the initial leader, and three followers, respectively are shown. As in the other plots, The subplots show the kinematic variables: (a) position, (b) velocity, (c) acceleration. All following vehicles react by reducing and then increasing the acceleration in order to maintain the platoon.

**Conclusions**

This paper has summarized and presented an architecture designed to evaluate truck platooning within traffic simulators. We have presented the main advantages and mechanisms to create the cross-platform architecture for simulating a truck-platooning functionality using Vissim and SymuVia simulation software. Initial evaluation results have shown appropriate behavior seen at operational level in a subset of maneuvers established for several trucks. Future research of this work involves the design of an assessment protocol to obtain specific key performance indicators for the efficiency and operational regulation of platoons in highways.

**References**

Calvert, S. C., W. J. Schakel, and B. van Arem. 2019. "Evaluation and Modelling of the Traffic Flow Effects of Truck Platooning." *Transportation Research Part C: Emerging Technologies* 105 (March): 1–22. https://doi.org/10.1016/j.trc.2019.05.019.

Calvert, S. C., W. J. Schakel, and J. W.C. van Lint. 2017. "Will Automated Vehicles Negatively Impact Traffic Flow?" *Journal of Advanced Transportation* 2017. https://doi.org/10.1155/2017/3082781.

Cicic, Mladen, and Karl Henrik Johansson. 2019a. "Energy-Optimal Platoon Catch-up in Moving Bottleneck Framework." *2019 18th European Control Conference, ECC 2019* 1: 3674–79. https://doi.org/10.23919/ECC.2019.8795754.

———. 2019b. "Stop-and-Go Wave Dissipation Using Accumulated Controlled Moving Bottlenecks in Multi-Class CTM Framework." *Proceedings of the IEEE Conference on Decision and Control* 2019-Decem: 3146–51. https://doi.org/10.1109/CDC40024.2019.9029216.

Duret, Aurelien, Meng Wang, and Andres Ladino. 2018. "A Hierarchical Approach for Splitting Truck Platoons near Network Discontinuities." *Transportation Research Procedia* 38 (July): 627–46. https://doi.org/10.1016/j.trpro.2019.05.033.

Jin, Li, Mladen Čičić, Karl H. Johansson, and Saurabh Amin. 2020. "Analysis and Design of Vehicle Platooning Operations on Mixed-Traffic Highways." *ArXiv*. https://doi.org/10.1109/tac.2020.3034871.

Kunze, Ralph, Richard Ramakers, Klaus Henning, and Sabina Jeschke. 2009. "Organization and Operation of Electronically Coupled Truck Platoons on German Motorways." *Lecture Notes in Computer Science (Including Subseries Lecture Notes in Artificial Intelligence and Lecture Notes in Bioinformatics)* 5928 LNAI (June): 135–46. https://doi.org/10.1007/978-3-642-10817-4_13.

Wang, Meng, Honghai Li, Jian Gao, Zichao Huang, Bin Li, and Bart Van Arem. 2018. "String Stability of Heterogeneous Platoons with Non-Connected Automated Vehicles." *IEEE Conference on Intelligent Transportation Systems, Proceedings, ITSC* 2018-March: 1–8. https://doi.org/10.1109/ITSC.2017.8317792.